\documentclass[10pt,twocolumn,shownopacs,preprintnumbers,amsmath,amssymb,showkeys]{revtex4}
\usepackage{graphicx}
\usepackage{dcolumn}
\usepackage{bm}
\begin{document}
\title{Transfer of single photon polarization states by two-channel continuous variable teleportation}
\author{Toshiki Ide}
\affiliation{%
Okayamaya Institute for Quantum Physics,\\
1-9-1 Kyoyama, Okayama City, Okayama, 700-0015, Japan
TEL: +81-86-256-3005 \quad FAX: +81-86-256-3580\\
e-mail: toshiki\_ide@pref.okayama.jp}
\author{Holger F. Hofmann}%
\affiliation{%
JST-CREST, Graduate School of Advanced Sciences of Matter, Hiroshima University,\\
Kagamiyama 1-3-1, Higashi Hiroshima 739-8530, Japan\\
TEL: +81-82-424-7652 \quad FAX: +81-82-424-7649\\
e-mail: hofmann@hiroshima-u.ac.jp
}%

\begin{abstract}
Superpositions of two orthogonal single-photon polarization states are commonly used as optical qubits. If such qubits are sent by continuous variable quantum teleportation, the modifications of the qubit states due to imperfect entanglement cause an increase in the average photon number of the output state. This effect can be interpreted as an accidental quantum cloning of the single photon input. We analyze the output statistics of the single photon teleportation and derive the transfer and cloning fidelities from the equations of the polarization qubit.
\end{abstract}


\maketitle

\section{Introduction}

Quantum teleportation is a method for transferring quantum states
to a remote location using quantum entanglement and
classical communication. 
The original proposal by Bennett et al. \cite
{Ben93} was based on a discrete set of basis states, and
its implementation for the two level system of single photon
polarization was experimentally realized by Bouwmeester
et al.\cite{Bou97}.
Vaidman \cite{Vai94} pointed out that it is
also possible to transfer quantum states using continuous variable
entanglement in its singular case. Later, continuous variable (CV)
teleportation was adapted to the physically realistic non-singular
case by Braunstein et al. \cite{Bra98} and the implementation
using entanglement obtained by optical squeezing was
experimentally realized by Furusawa et al. \cite{Fur98}.

The transfer of two level systems such as the single photon
polarization state teleported in ref. \cite{Bou97} is of
particular significance for quantum information because it
represents a qubit, the smallest unit of quantum information.
It may therefore be of interest to consider the transfer of
photon polarization qubits using the CV teleportation scheme
reported in ref. \cite{Fur98}.
As we discussed previously \cite{Ide01},
this can be achieved by separately applying the standard CV
teleportation protocol to a pair of orthogonally polarized modes.
Our discussion in ref.\cite{Ide01} showed that this method can
result in different output photon numbers in the teleported state.
The teleported photon can either be lost, resulting in a vacuum
output component, or it can be multiplied, resulting in an $N$-photon
output state. The latter effect can be interpreted as an
accidental quantum cloning procedure applied to the single photon
input.

In our previous work, we derived the transfer operator for
CV teleportation \cite{Hof00} and applied it to single photon
teleportation problems \cite{Ide01}. In the following, we analyze
the output statistics for the qubit teleportation in terms of
the transfer and cloning fidelities for different photon number
outputs. It is shown that the cloning fidelities approach their
optimal values for highly entangled states.

\section{Two-mode transfer operator formalism}

\begin{figure}[htbp]
\begin{picture}(200,150)

\put(80,20){\framebox(40,20){\large OPA}}

\put(80,42.5){\line(-1,1){22.5}}
\put(77.5,40){\line(-1,1){22.5}}
\put(55,65){\line(0,-1){5}}
\put(55,65){\line(1,0){5}}
\put(67.5,57.5){\makebox(10,10){\large $R$}}

\put(120,42.5){\line(1,1){22.5}}
\put(122.5,40){\line(1,1){22.5}}
\put(145,65){\line(0,-1){5}}
\put(145,65){\line(-1,0){5}}
\put(122.5,57.5){\makebox(10,10){\large $B$}}

\put(20,42.5){\line(1,1){22.5}}
\put(22.5,40){\line(1,1){22.5}}
\put(45,65){\line(0,-1){5}}
\put(45,65){\line(-1,0){5}}
\put(22.5,57.5){\makebox(10,10){\large $A$}}

\put(5,33){\makebox(20,6){Input}}
\put(5,23){\makebox(20,6){field}}

\put(50,55){\line(0,1){30}}
\put(40,45){\makebox(20,6){Beam}} 
\put(40,35){\makebox(20,6){splitter}}

\put(45,77.5){\line(-1,1){22.5}}
\put(42.5,75){\line(-1,1){22.5}}
\put(20,100){\line(0,-1){5}}
\put(20,100){\line(1,0){5}}
\put(15,102){\makebox(10,10){\large $x_{H-/V-}$}}

\put(55,77.5){\line(1,1){22.5}}
\put(57.5,75){\line(1,1){22.5}}
\put(80,100){\line(0,-1){5}}
\put(80,100){\line(-1,0){5}}
\put(75,102){\makebox(10,10){\large $y_{H+/V+}$}}

\put(5,115){\framebox(90,40){}}
\put(30,140){\makebox(40,10){Measurement of}}
\put(30,130){\makebox(40,10){$\beta_{H}=x_{H-}+i y_{H+}$}}
\put(30,120){\makebox(40,10){$\beta_{V}=x_{V-}+i y_{V+}$}}

\bezier{200}(95,135)(125,130)(155,90)
\put(155,90){\line(0,1){10}}
\put(155,90){\line(-2,1){10}}

\put(127.5,69.5){\framebox(60,20){$\hat{D}_{HV}(\beta_H,\beta_V)$}}

\put(170,92.5){\line(1,1){12.5}}
\put(172.5,90){\line(1,1){12.5}}
\put(185,105){\line(0,-1){5}}
\put(185,105){\line(-1,0){5}}

\put(175,117.5){\makebox(20,6){Output}}
\put(175,107.5){\makebox(20,6){field}}
\end{picture}
\caption{Schematic representation of the two-mode quantum teleportation
setup. The entangled state is generated by four mode squeezing in an optical parametric amplifier (OPA). Four separate homodyne detection
measurements are used to obtain the polarization components of the
complex displacement amplitudes $\beta_H$ and $\beta_V$.}
\label{setup}
\end{figure}
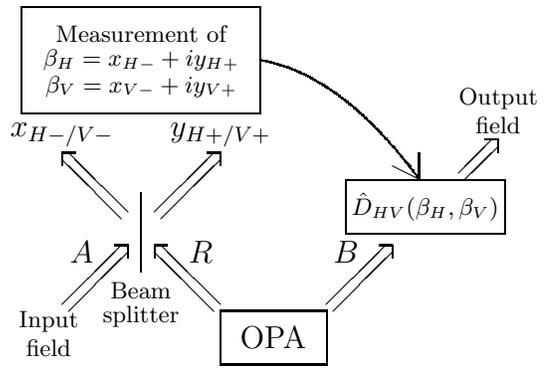

Fig.\ref{setup} shows the schematic setup of the CV quantum
teleportation for the two polarization modes of the input field $A$.
Each local light field is now described by two polarization modes.
An optical parametric amplifier (OPA) generates the
squeezed state entanglement between the $R$ and $B$ modes
in both the horizontal ($H$) and the vertical ($V$) polarizations,
resulting in a pair of beams equally
entangled in every possible polarization direction.
This four-mode squeezed entangled state can be written as
\begin{eqnarray}
&&\mid  \mbox{EPR}(q) \rangle_{R,B} = \nonumber \\
&& (1-q^2) \sum_{n_H,n_V=0}^{\infty} q^{(n_H+n_V)} \mid
n_H;n_V \rangle_{R} \otimes \mid n_H;n_V \rangle_{B},\nonumber \\
\end{eqnarray}
where $R$ is the mode used by Alice (sender) as a quantum reference
in the joint measurement of $A$ and $R$, and $B$ is the output
mode on Bob's side (receiver). The variable $q$ defines the
degree of squeezing, and thus the amount of entanglement of the
state. 
It may be worth noting that this entangled state can be
written as a product state of two entangled states for
any pair of orthogonal polarizations.

It is now possible to derive a transfer operator
$\hat{T}_{\mbox{pol.}} (\beta_H, \beta_V)$ for the two mode
teleportation, so that the output state in $B$ is given by
\cite{Hof00}
\begin{equation}
\mid \psi_{\mbox{out}}(\beta_H,\beta_V) \rangle_{HV} =
\hat{T}_{\mbox{pol.}} (\beta_H, \beta_V) \mid \psi_{in}
\rangle_{HV}.
\label{trans}
\end{equation}
As has been shown in ref.\cite{Ide01}, this transfer operator
can be obtained by a product of two single mode transfer operators.
Using the results of ref.\cite{Hof00,Hof01}, the transfer operator
$\hat{T}_{\mbox{pol.}}$ can therefore be expressed using the
Fock states $\mid n_H;n_V \rangle_{HV}$ and the displacement
operators $\hat{D}_{HV}(\beta_H,\beta_V)$ can therefore be expressed in its diagonal form by the displaced Fock states 
$\hat{D}_{HV}(\beta_H,\beta_V)\mid ~ n_H;n_V \rangle_{HV}$,
\begin{eqnarray}
\lefteqn{
\hat{T}_{\mbox{pol.}}(\beta_H, \beta_V)=\hat{T}_{H,q}(\beta_H) \otimes
\hat{T}_{V,q}(\beta_H)} \nonumber \\
&=& \frac{1-q^2}{\pi} \sum_{n_H=0}^{\infty}
\sum_{n_V=0}^{\infty} q^{n_H+n_V}
\nonumber \\ &\times& 
\hat{D}_{HV}(\beta_H,\beta_V)
\mid n_H;n_V \rangle \langle n_H;n_V \mid
\hat{D}_{HV}(-\beta_H,-\beta_V).
\nonumber \\
\label{newtrans}
\end{eqnarray}
In principle we can decompose the polarization transfer operator $\hat{T}_
{\mbox{pol.}}$ into any two orthogonal modes (e.g. right and left
circular polarization mode, $\pm 45^{\circ}$ polarization mode).
The choice of horizontal ($H$) and vertical ($V$) polarization
is made to simplify the labeling.

We can now apply this transfer operator to the specific case of
single photon inputs. For an unknown polarization, the
input state is described by a superposition of the two
basis states, $c_H \mid 1;0 \rangle + c_V \mid 0;1 \rangle$.
However, as noted above, the choice of the polarization basis
is arbitrary. Therefore, all polarization states are teleported
equally well, and it is sufficient to derive the output statistics
for the horizontally polarized input $\mid 1;0 \rangle$.
As explained in more detail in ref.\cite{Ide01}, the results
for the fidelities and the output photon number distributions
will be the same for any other input polarization.

The output state for the teleportation of a single horizontally
polarized qubit reads \cite{Ide01}
\begin{eqnarray}
\mid\psi_{\mbox{out}}\rangle_{HV} &=&
\hat{T}_{H,q}(\beta_H)\mid 1 \rangle \otimes
\hat{T}_{V,q}(\beta_V) \mid 0 \rangle 
\nonumber \\
&=&\frac{1-q^2}{\pi}
e^{-(1-q^2) (|\beta_H|^2+|\beta_V|^2)/2}\nonumber\\
&&\times\hat{D}_{HV}((1-q)\beta_H, (1-q)\beta_V)\nonumber\\
&&
\left ( (1-q^2) \beta_H^{\ast} \mid 0;0 \rangle +
q \mid 1;0 \rangle \right).
\label{10out}
\end{eqnarray}
Since this state is essentially a product state of the
vacuum teleportation in the vertical polarized component
and the one photon teleportation in the horizontally
polarized component, we can now determine the photon
statistics of the output using the results derived in
ref.\cite{Ide01}.

\section{Continuous variable teleportation of photon polarization}

The average numbers of horizontally and vertically polarized
photons in the output of the teleportation can be determined
by averaging the results given in eq. (\ref{10out}) over all
measurement values of $\beta_H$ and $\beta_V$ \cite{Ide01}.
The results read
\begin{eqnarray}
\langle n_H \rangle &=& \frac{2}{1+q},
\nonumber \\
\langle n_V \rangle &=& \frac{1-q}{1+q}.
\end{eqnarray}
The effect of CV teleportation on the average photon numbers is
simply given by an addition of $(1-q)/(1+q)$ photons to each
channel. This additional intensity originates from the ``quantum duty"
paid for non-maximal entanglement, as explained by Braunstein et
al. in the original proposal of optical CV teleportation
\cite{Bra98}. Since the additional photons are unpolarized, the
polarization fidelity of the output photons is reduced to
\begin{equation}
F_{\mbox{av.}}=\frac{\langle n_H \rangle}{\langle n_H \rangle
+ \langle n_V\rangle}=\frac{2}{3-q}.
\end{equation}
As expected, this value increases from the classical teleportation
limit of $2/3$ at no entanglement ($q=0$) to a perfect fidelity of
one at maximal entanglement ($q=1$).

The effect of CV teleportation on the average polarization
fidelity $F_{\mbox{av.}}$ is therefore easily explained by the
``quantum duty" according to Braunstein et al.\cite{Bra98}.
However, this result does not show how the teleportation errors
and the output photon number are correlated.
We can now apply our more detailed results from ref.\cite{Ide01}
to derive this correlation and to identify the origin of the
polarization errors.
As shown in ref.\cite{Ide01}, the output photon number distribution
of the single-mode teleportation of a single photon input
is
\begin{eqnarray}
p^1_{q}(n)&=&\int d^2\beta | \langle n \mid \hat{T}_{q}(\beta)\mid 1
\rangle|^2 \nonumber \\
&=& \frac{1+q}{2} \left ( \frac{1-q}{2} \right )^{n+1}
\left ( 1+ n \left ( \frac{1+q}{1-q} \right )^2 \right ).
\label{p1}
\end{eqnarray}
In the case of a zero photon input, the output photon number
distribution corresponds to a thermal distribution with
\begin{eqnarray}
p^0_{q}(n)&=&\int d^2\beta | \langle n \mid \hat{T}_{q}(\beta)\mid 0
\rangle|^2 \nonumber \\
&=& \frac{1+q}{2} \left ( \frac{1-q}{2} \right )^{n}.
\label{p0}
\end{eqnarray}
Since the teleportation of the horizontally polarized photon
given by eq.(\ref{10out}) can be written as a product of a
single photon polarization in $H$ and a vacuum teleportation
in $V$, the output statistics of the two polarizations
is given by the product of the probabilities given above.
The joint probability thus reads
\begin{eqnarray}
&&p^{1,0}(n_H,N-n_H)\nonumber\\
&=&\left ( \frac{1+q}{2} \right )^2 \left ( \frac{1-q}{2} \right )^{N+1}
\left ( 1+ n_H \left ( \frac{1+q}{1-q} \right )^2 \right ),
\label{jointp}
\end{eqnarray}
where $N=n_H+n_V$ is the total photon number in the output.

It is now possible to determine the polarization fidelities
for a specific number of output photons. For example, it
is possible to post-select the cases where only a single
photon is found in the output. As previously mentioned
in \cite{Ide01}, the conditional polarization fidelity
$F_1$ for this post-selected output is given by
\begin{eqnarray}
F_{\mbox{1}}=
\frac{p^{1,0}(1,0)}{p^{1,0}(1,0)+p^{1,0}(0,1)}=\frac{2(1+q^2)}
{2(1+q^2)+(1-q)^2}.
\end{eqnarray}
\begin{figure}[htbp]
\begin{picture}(220,150)
\put(200,145){\makebox(40,10){\Large $q$}}
\put(60,100){\makebox(40,10){\Large $F_{\mbox{1}}$}}
\put(120,55){\makebox(40,10){\Large $F_{\mbox{av.}}$}}
\put(-10,145){\makebox(40,10){\Large $F$}}
\includegraphics[width=8cm]{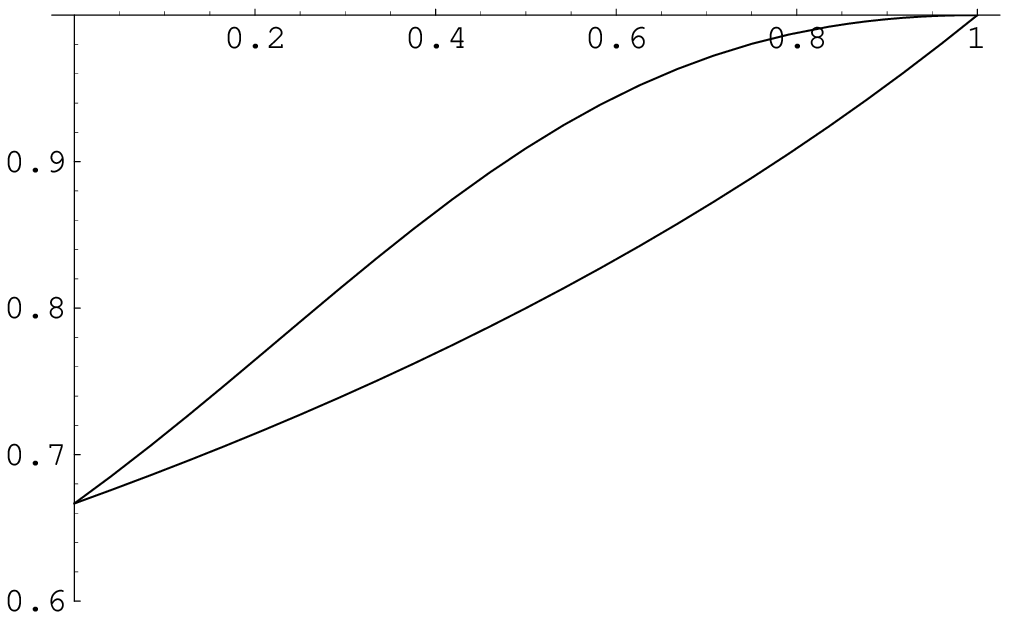}
\end{picture}
\caption{Comparison of the average polarization fidelity
$F_{\mbox{av.}}$ and the post-selected fidelity $F_{\mbox{1}}$
of the single photon output for variable entanglement
parameter $q$.}
\label{avepos}
\end{figure}

Fig.\ref{avepos} shows the comparison between this post-selected
fidelity and the average polarization fidelity $F_{\mbox{av.}}$.
It can be seen that $F_{\mbox{1}}$ is higher than
$F_{\mbox{av.}}$ for all values of the entanglement parameter $q$.
We can therefore conclude that the errors that decrease the
fidelity to $F_{\mbox{av.}}$ originate from multi photon outputs
($N>1$). As we will discuss in the next section, this result
can be interpreted in terms of the limits on quantum cloning
imposed on the cloning fidelities $F_N$ of the multi photon
outputs.

\section{Cloning fidelity}

Since continuous variable teleportation does not preserve the
input photon number, it is possible to detect more than one
output photon in the teleportation of a single input photon.
Specifically, the probability $P_q(N)$ of obtaining $N$ photons
in the output is given by
\begin{eqnarray}
&&P_q(N)=\sum^N_{n_H=0} p^{1,0}(n_H,N-n_H)\nonumber\\
&=&(N+1)\left ( \frac{1+q}{2} \right )^2 \left ( \frac{1-q}{2} \right )^{N+
1} \left ( 1+ \frac{N}{2} \left ( \frac{1+q}{1-q} \right )^2 \right ).
\nonumber\\
\label{pneq}
\end{eqnarray}
\begin{figure}[htbp]
\begin{picture}(220,150)
\put(200,-10){\makebox(40,10){\Large $q$}}
\put(-10,145){\makebox(40,10){\Large $P_q(N)$}}
\includegraphics[width=8cm]{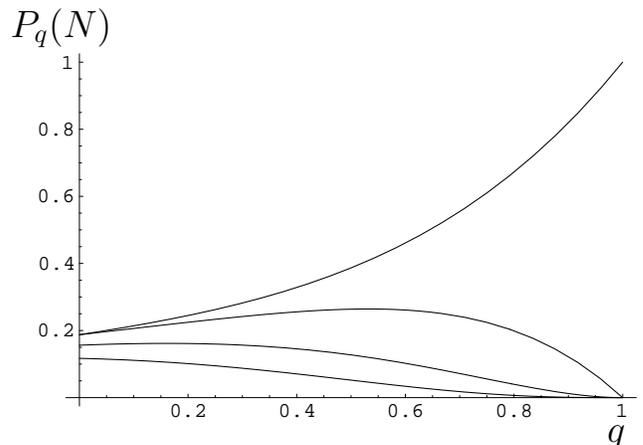}
\end{picture}
\caption{Probability of obtaining $N$-photon output as a function
entanglement parameter $q$. From top
to bottom, the curves show $P_q(1), P_q(2), P_q(3)$ and $P_q(4)$.}
\label{pn}
\end{figure}

Fig.\ref{pn} shows the probabilities for $N=1$, $N=2$, $N=3$, and
$N=4$ output photons. For finite entanglement ($q<1$), there is
always a non-vanishing chance of generating additional photons.
However, the probabilities of obtaining high photon numbers $N$
decrease rapidly with increasing entanglement $q$, consistent
with the decrease in the average output photon number due
to the reduced ``quantum duty".

Since the output photons are indistinguishable, the
polarization state of the input photon is transferred
equally to all of them.
It is therefore possible to consider all of
the $N$ output photons as clones of the original input photon.
The conditional probability of finding any one of the $N$ photons
in the same polarization as the input state is therefore
equal to the cloning fidelity $F_N$.
This fidelity can be determined from the joint probabilities
given in eq. (\ref{jointp}). It reads
\begin{eqnarray}
F_N =
\left( \frac{2}{3}+ \frac{\left(\displaystyle\frac{1+q}{1-q}\right)^2 -1}
{3N\left( \displaystyle\frac{1+q}{1-q}\right)^2 +6}\right).
\label{cloning}
\end{eqnarray}

\begin{figure}[htbp]
\begin{picture}(220,150)
\put(200,135){\makebox(40,10){\Large $q$}}
\put(-10,135){\makebox(40,10){\Large $F_N$}}
\includegraphics[width=8cm]{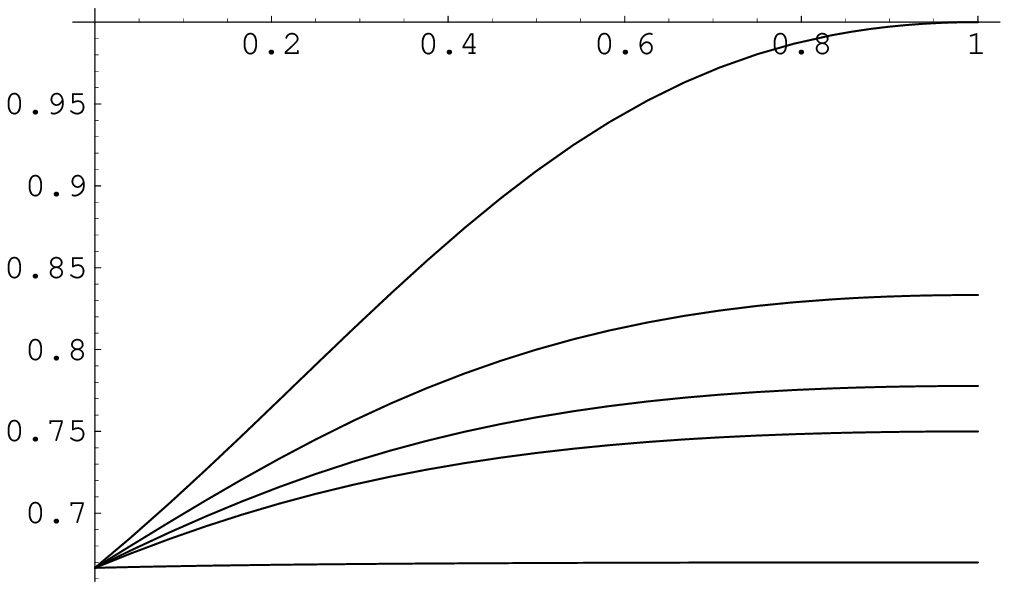}
\end{picture}
\caption{Cloning fidelities $F_N$ for different output photon
numbers $N$. From top to bottom, the curves show
$F_1, F_2, F_3, F_4$ and $F_{100}$.}
\label{cfid2}
\end{figure}
Fig.\ref{cfid2} shows the single photon fidelity $F_1$ and the
cloning fidelities for two to four output photons. 
In addition, the high $N$ limit is indicated by the result
for $N=100$, which is basically indistinguishable from a
flat line at $F_\infty=2/3$.
As eq.(\ref{cloning}) shows, all cloning fidelities
$F_N$ are equal to $2/3$ for classical teleportation at
$q=0$. The fidelities increase with $q$ and reach their maximal
value at $q=1$. However, the limit for $q\to 1$ cannot be one
for $N > 1$, because it would contradict the no-cloning
theorem \cite{Woo96}. Therefore, the upper limit depends on
$N$ according to
\begin{eqnarray}
\frac{2}{3} \leq F_N \leq \frac{2N+1}{3N}.
\label{cloningeq}
\end{eqnarray}
This upper limit is equal to the maximal cloning fidelity
possible for a cloning protocol producing $N$ output photons
from one input photon. For an infinite number of clones,
this limit is equal to the teleportation fidelity of
classical teleportation, as indicated by the flat
line for $F_{100}$ in fig. \ref{cfid2}.

\section{Conclusions and Outlook}

We have analyzed the CV teleportation of a single photon polarization
state by applying the transfer operator formalism to two orthogonally
polarized teleportation channels.
The modifications of the output states due to imperfect entanglement
result in an increase in the average photon number in the output
due to the ``quantum duty" paid for the use of non-maximal
entanglement.
The average polarization fidelity of all output photons is
therefore reduced by the random polarization of this ``quantum duty".
However, our results also show that the conditional polarization
fidelity of the single photon transfer is much higher than the
average fidelity, indicating that the polarization errors are
greater in the multi-photon outputs.
Since the transfer of a single photon to a multi-photon output
corresponds to a quantum cloning process, this observation can be
understood as a result of the no cloning theorem.
Indeed, our detailed analysis shows that the cloning fidelities
for each $N$ photon output approaches the maximal possible value
for quantum cloning protocols as the entanglement parameter $q$
approaches one. A significant part of the polarization errors
in the output photons thus originates from the cloning
errors associated with the accidental generation of additional
output photons.

From the quantum communication point of view, it seems to be
quite remarkable that CV teleportation can produce nearly
optimally cloned qubits as an accidental side effect of
qubit teleportation. In this context, the relation with
intentional telecloning protocols \cite{Mur99} may be of further interest.

\end{document}